\documentclass[slac_one]{revtex4}
\usepackage{graphicx}
\usepackage{fancyhdr}
\usepackage{axodraw}
\usepackage{amsmath}
\begin{document}

%Title of paper
\title{{\small{Hadron Collider Physics Symposium (HCP2008),
Galena, Illinois, USA}}\\ %% Please keep this conference title here
\vspace{12pt}
Experimental Results on Diffraction} %% Paper title goes here

% Repeat the \author .. \affiliation  etc. as needed
%
% \affiliation command applies to all authors since the last
% \affiliation command. The \affiliation command should follow the
% other information

\author{P.\@ Van Mechelen}
\affiliation{Universiteit Antwerpen, Belgium}

\begin{abstract}
Results on diffractive scattering observed at HERA and at the TEVATRON are reviewed.  This includes the extraction of diffractive parton density functions and determination of the rapidity gap survival probability at HERA and  the observation of central exclusive production of final states at the TEVATRON. Finally, preparations to observe diffractive signals at the LHC are discussed.  
\end{abstract}

%\maketitle must follow title, authors, abstract
\maketitle

\thispagestyle{fancy}

% body of paper here - Use proper section commands
% References should be done using the \cite, \ref, and \label commands
% Put \label in argument of \section for cross-referencing
%\section{\label{}}

\section{DIFFRACTIVE PROCESSES AND KINEMATICS}

\subsection{Proton-Proton Diffraction}

In single diffractive dissociation (SDD), $pp \to pX$, one of the protons survives the interaction while the other dissociates in a hadronic system with invariant mass $M_X$, separated from the first proton by a large rapidity interval devoid of particles.  In the presence of a hard scale, such interactions may be regarded as the result of the exchange of a colourless object with vacuum quantum numbers (e.g.\@ a pomeron) consisting of quarks and gluons. One further defines $\xi = 1 - \frac{P'_L}{P_L}$, the fractional longitudinal momentum loss of the surviving proton and $t = (P - P')^2$, the squared four-momentum exchange at the proton vertex, with $P$ and $P'$ the four-momenta of the initial and final state proton, respectively, measured in the initial state centre-of-mass frame.

Double pomeron exchange (DPE), $pp \to pXp$, is the process where both protons survive the interaction, whilst a central hadronic system with invariant mass $M_X$ is produced through the fusion of two colourless objects (often assumed to be pomerons).  One consequently defines $\xi_1$, $\xi_2$, $t_1$ and $t_2$ as above, with the indices referring to one of both proton vertices.  In hard central exclusive production (CEP), the central hadronic system boasts a hard scale (transverse momentum, invariant mass, \ldots) with no soft remnants present in the final state $X$.

\subsection{Electron-Proton Diffraction}

\begin{figure}[htb]
\begin{picture}(400,135)(0,5)
\SetWidth{1}
\Text(80,5)[]{(a)}
\Line(10,20)(150,20) \Text(10,22)[b]{$p (P)$} \Text(150,22)[b]{$p' (P')$}
\Line(78,20)(78,60)
\Line(82,20)(82,60) \Text(76,40)[r]{$I\!\!P$}
\Line(80,60)(130,60) \Text(132,60)[l]{$\Big\}X (X)$}
\Line(80,60)(130,68)
\Line(80,60)(130,52)
\Photon(80,60)(50,90){3}{4} \Text(65,73)[tr]{$\gamma^\ast$}
\Line(10,90)(50,90) \Text(10,92)[b]{$e (k)$}
\Line(50,90)(120,120) \Text(118,120)[br]{$e' (k')$}
\GCirc(80,20){4}{1}
\GCirc(80,60){4}{1}
\Text(320,5)[]{(b)}
\Line(250,20)(390,20)  \Text(250,22)[b]{$p (P)$} \Text(390,22)[b]{$p' (P')$}
\Line(318,20)(318,50)
\Line(322,20)(322,50) \Text(316,35)[r]{$I\!\!P$}
\Line(320,50)(350,50)
\Line(320,50)(350,55)
\Line(320,50)(350,45)
\Gluon(320,50)(340,70){-3}{3} \Text(328,60)[br]{$(v)$}
\Gluon(320,90)(340,70){3}{3} \Text(328,80)[tr]{$(u)$}
\Line(340,70)(370,78)
\Line(340,70)(370,62)  %\Text(372,70)[l]{$X$}
\Line(320,90)(350,90)
\Line(320,90)(350,95)
\Line(320,90)(350,85)
\Photon(320,90)(290,110){3}{4} \Text(300,98)[tr]{$\gamma$}
\Line(250,110)(290,110) \Text(250,112)[b]{$e (k)$}
\Line(290,110)(360,140) \Text(358,140)[br]{$e' (k')$}
\GCirc(320,20){4}{1}
\GCirc(320,50){4}{1}
\GCirc(340,70){4}{1}
\GCirc(320,90){4}{1}
\end{picture}
\caption{(a)~Diagram representing diffractive deep-inelastic scattering.  (b)~Diagram representing diffractive photoproduction. The four-momenta of the particles involved are given in parentheses.} \label{fig:epdiff}
\end{figure}
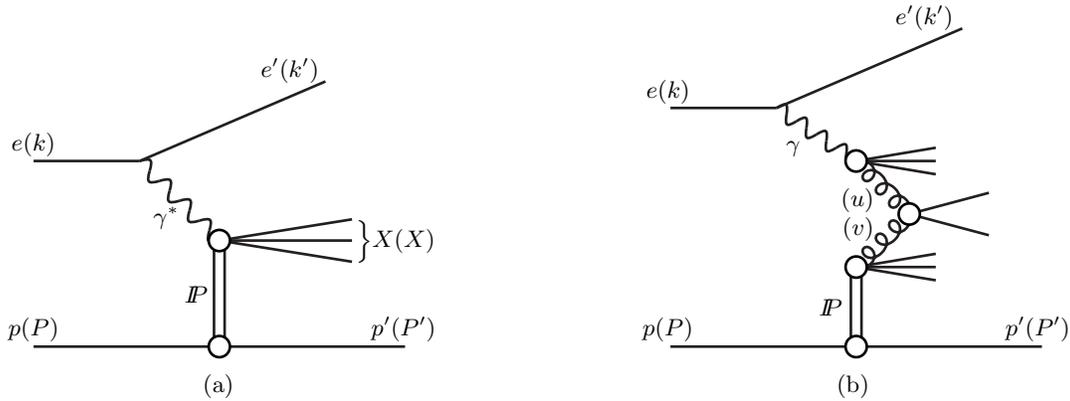

Diffractive deep-inelastic scattering (DDIS), $ep \to e \gamma^\ast p \to eXp$ occurs through the fusion of a virtual photon emitted by the electron and a  colourless object exchanged by the proton (see Fig.\@~\ref{fig:epdiff}a).  Besides the usual deep-inelastic scattering variables, the photon virtuality $Q^2 = -q^2 = - (k-k')^2$ and Bjorken-$x = Q^2/2 q\cdot P$, one defines $t = (P-P')^2$ as the squared four-momentum transfer at the proton vertex, $M_X$ as the invariant mass of the photon dissociation system, $x_{I\!\!P} = q \cdot (P-P')/ q \cdot P$ as the fractional momentum loss of the proton ($x_{I\!\!P}$ is equivalent to the variable $\xi$ used in proton-proton diffraction), and $\beta=x/x_{I\!\!P}$ as the momentum fraction of the pomeron carried by the struck quark.  

In diffractive photoproduction (DPHP), $ep \to e \gamma p \to eXp$, a quasi-real photon emitted by the electron interacts diffractively with the proton (see Fig.\@~\ref{fig:epdiff}b) to produce a central hadronic system $X$.  If this system has a hard scale, one may define $x_\gamma = P \cdot u / P \cdot q$ as the fractional momentum from the photon entering the hard interaction and $z_{I\!\!P} = q \cdot v / q \cdot (P-P')$ as the fractional momentum from the colourless exchange transferred to the hard interaction.  The four-momenta used in the above formulae are defined in the figure.

\section{MEASURING DIFFRACTIVE PARTON DENSITY FUNCTIONS}

\subsection{Experimental Selection and Cross Section Measurement}

The HERA experiments use different methods for selecting diffractive interactions: requiring the presence of a large rapidity gap, exploiting the shape of the $M_X$ distribution or using direct proton tagging.  

In the rapidity gap method, one requires a large interval in rapidity devoid of particles.  This interval typically spans the range $3.3 < \eta < 7.5$ in the laboratory frame.  The kinematics of the event is reconstructed from the photon dissociation system $X$. The four-momentum squared $t$ is not measured but integrated over and, in the case of the H1 detector, the experimental selection ensures that, if the proton dissociates, this dissociation system has an invariant mass $M_Y < 1.6 {\rm\ GeV}$.  

\begin{figure}[htb]
\includegraphics[angle=270,width=0.70\textwidth]{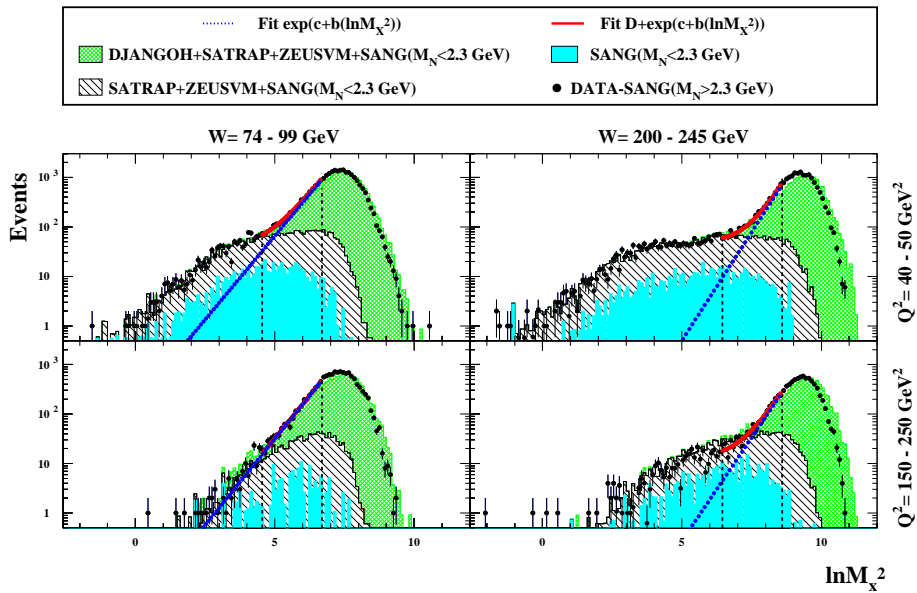}
\caption{The $\ln M_X^2$ distribution as obtained by the ZEUS experiment. Points represent the data, while the coloured histograms show the constributions to the total event yield of diffractive (grey hatched) and non-diffractive (green hatched) DIS as described by Monte Carlo models~\cite{bib:mxzeus}.} \label{fig:mxmethod}
\end{figure}

Another possibility is to extract a diffractive event sample from a fit to the $M_X$ distribution.  As can be seen in Fig.\@~\ref{fig:mxmethod}~\cite{bib:mxzeus}, the non-diffractive background falls off exponentially towards low $M_X$ and a fit of the form $D+\exp(c+b \ln M_X^2)$ will yield the diffractive contribution $D$.  As in the rapidity gap method, the kinematics of the event is measured from the $X$ system.  Again, one integrates over $t$ and, in the case of the ZEUS detector, the mass of the proton dissocation system is limited to $M_Y < 2.3 {\rm\ GeV}$

The most straightforward method is direct proton tagging with forward proton detectors.  In this case, a pure single diffractive event sample is obtained without any contamination by proton dissociation events and a direct reconstruction of $t$ is possible through the measurement of the proton four-momentum.

\begin{figure}[htb]
\includegraphics[width=0.8\textwidth]{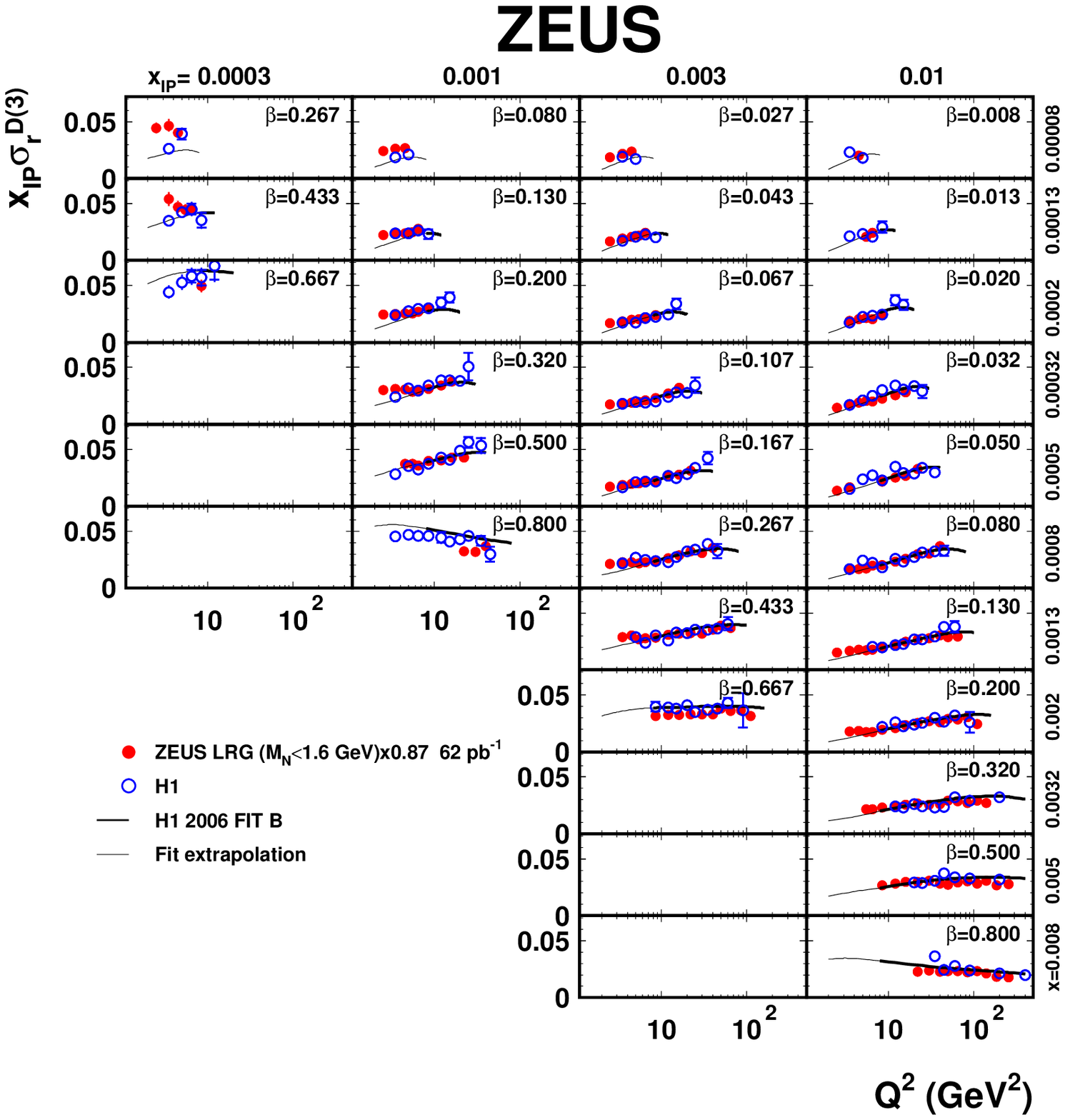}
\caption{The reduced cross section $\sigma_r^{D(3)} = \frac{d\sigma^D}{dx_{I\!\!P}\, dx\, dQ^2} / \frac{4\pi\alpha^2}{xQ^4}\left(1 - y + \frac{y^2}{2}\right)$ is plotted against $Q^2$ in bins of $x$ and  $x_{I\!\!P}$. H1 and ZEUS data are compared to the H1 2006 Fit B (see futher in the text).  A normalisation difference between ZEUS and H1 data is not shown (the ZEUS data points are scaled down by 13\%).} \label{fig:sigmaddis}
\end{figure}

Figure~\ref{fig:sigmaddis} shows, as an example, the DDIS cross section obtained with the large rapidity gap method by the ZEUS and H1 experiments~\cite{bib:h1zeuslrg}.  Good agreement, within experimental uncertainties, is obtained between both experiments.  A remaining normalisation difference of 13\% is covered by the uncertainty on the proton dissociation correction (8\%) and the relative normalisation uncertainty (7\%).  Results obtained with different selection methods also agree well.

\subsection{QCD and Vertex Factorisation}

In the QCD analysis of DDIS one assumes two different forms of factorisation.  QCD hard scattering factorisation has been theoretically proven to hold in DDIS \cite{bib:collins} and separates the partonic hard scattering cross section $\sigma^{ei}$, for the interaction between the electron and a quark $i$ out of the proton, from a so-called diffractive parton density function (DPDF) $f_i^D$, which describes the probability to find a quark inside the proton under the condition that the proton survives the interaction with kinematics described by $x_{I\!\!P}$ and $t$:
\begin{equation}
\sigma^{ep \to eXp} (x, Q^2, x_{I\!\!P}, t) = \sum_i f_i^D(x,Q^2,x_{I\!\!P},t) \cdot \sigma^{ei}(x,Q^2). 
\end{equation}

Proton vertex (or Regge) factorisation, on the other hand, is only approximately satisfied.  Nevertheless, it can be used successfully in the parametrisation of the DDIS cross section.  This factorisation assumption expresses the DPDF as a superposition of pomeron and reggeon terms, separating the flux factors $f_{I\!\!P/p}$ and $f_{I\!\!R/p}$ of pomerons and reggeons in the proton from their partonic structure $f_i^{I\!\!P}$ and $f_i^{I\!\!R}$:
\begin{equation}
f_i^D(x,Q^2,x_{I\!\!P},t) = f_{I\!\!P/p}(x_{I\!\!P},t) \cdot f_i^{I\!\!P}(\beta=\frac{x}{x_{I\!\!P}}, Q^2) + n_{I\!\!R} f_{I\!\!R/p}(x_{I\!\!P},t) \cdot f_i^{I\!\!R}(\beta=\frac{x}{x_{I\!\!P}}, Q^2). 
\end{equation}
Here, $n_{I\!\!R}$ is a factor describing the relative normalisation of reggeon to pomeron fluxes.  The fluxes themselves are obtained from a parameterisation inspired by Regge Theory where the $x_{I\!\!P}$ dependence of the pomeron flux is governed by the parameter $\alpha_{I\!\!P}(0)$.

\subsection{From Cross Sections to Diffractive Parton Density Functions}

A NLO QCD fit can be performed yielding values for $\alpha_{I\!\!P}(0)$, $n_{I\!\!R}$ and a polynomial for the quark and gluon densities of the pomeron at a fixed starting scale $Q_0^2$.  Usually, the reggeon flux is  fixed and its parton density is taken to be equal to that of the pion.  

\begin{figure}[htb]
\includegraphics[width=0.5\textwidth]{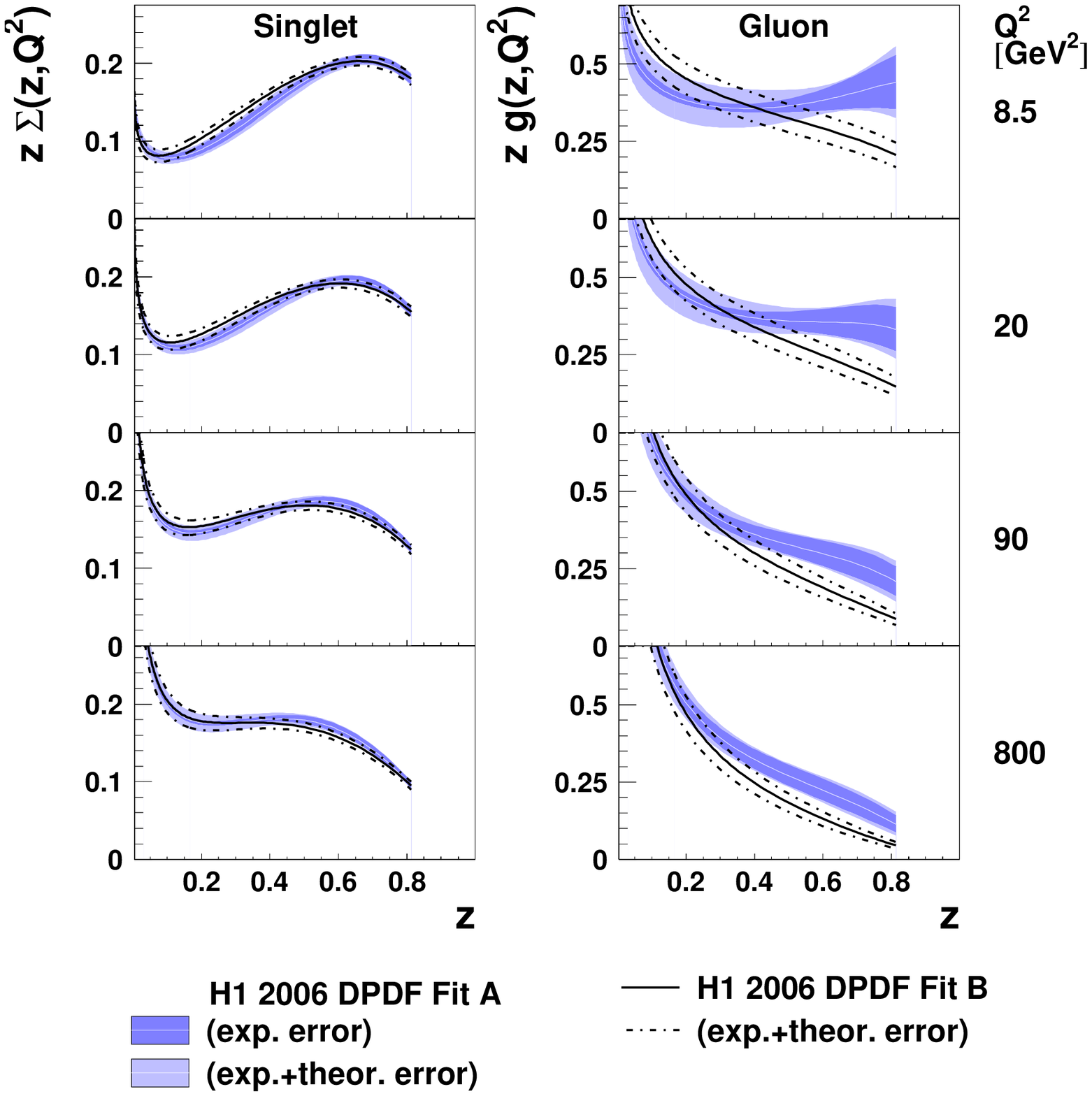}
\caption{The quark (singlet) and gluon densities as obtained in a NLO QCD fit are shown as function of fractional momentum $z$ at different scales $Q^2$.  Two fits are obtained based on different parametrisations of the gluon density at the starting scale $Q_0^2$.} \label{fig:h1fit}
\end{figure}

The H1 collaboration obtained two fits (labelled $A$ and $B$) using different polynomial forms for the gluon distribution at the starting scale (see Fig.\@~\ref{fig:h1fit}) \cite{bib:ddis_dpdf}.  Both have similar good $\chi^2$ values of 158/183 d.o.f.\@ and 164/184 d.o.f., respectively.  The quark distributions are found to be very stable in both fits, while the gluon distributions agree at low values of $z$ but vary at high $z$. 

\begin{figure}[htb]
\includegraphics[width=0.3\textwidth]{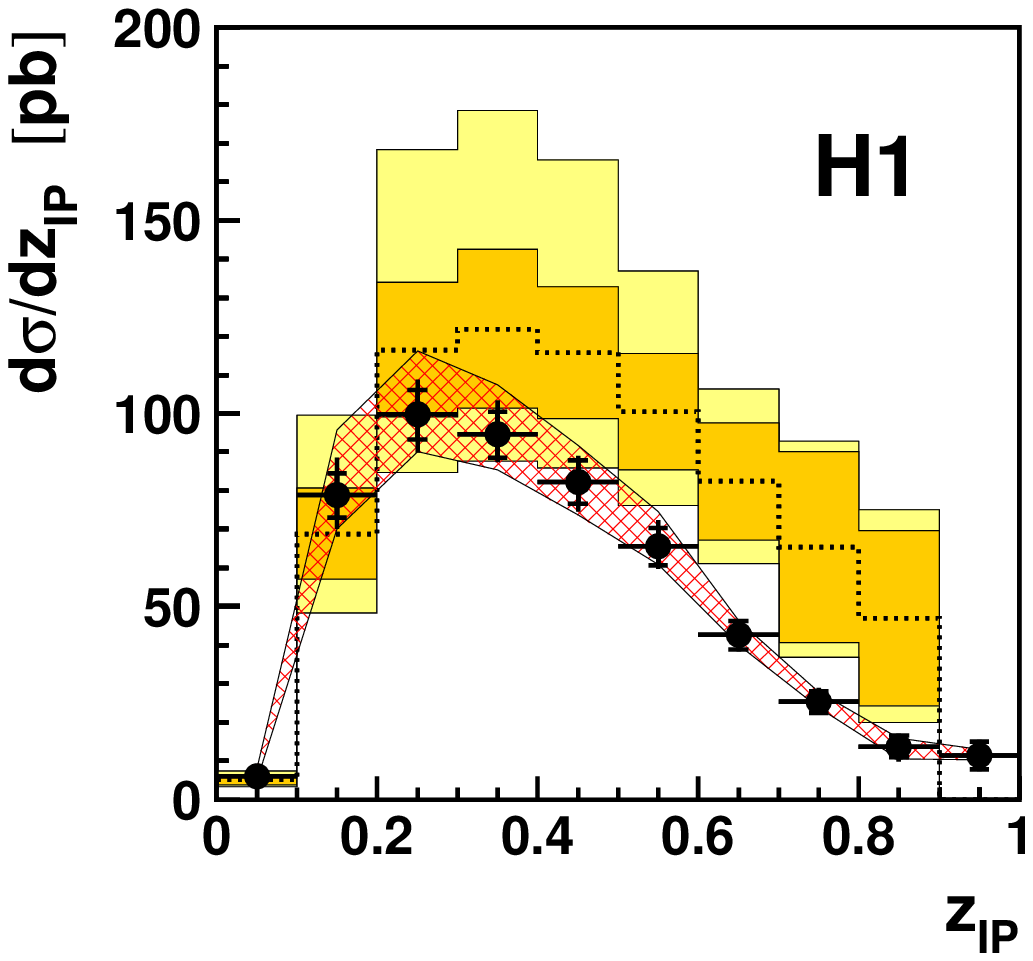}\hspace{1cm}\includegraphics[width=0.3\textwidth]{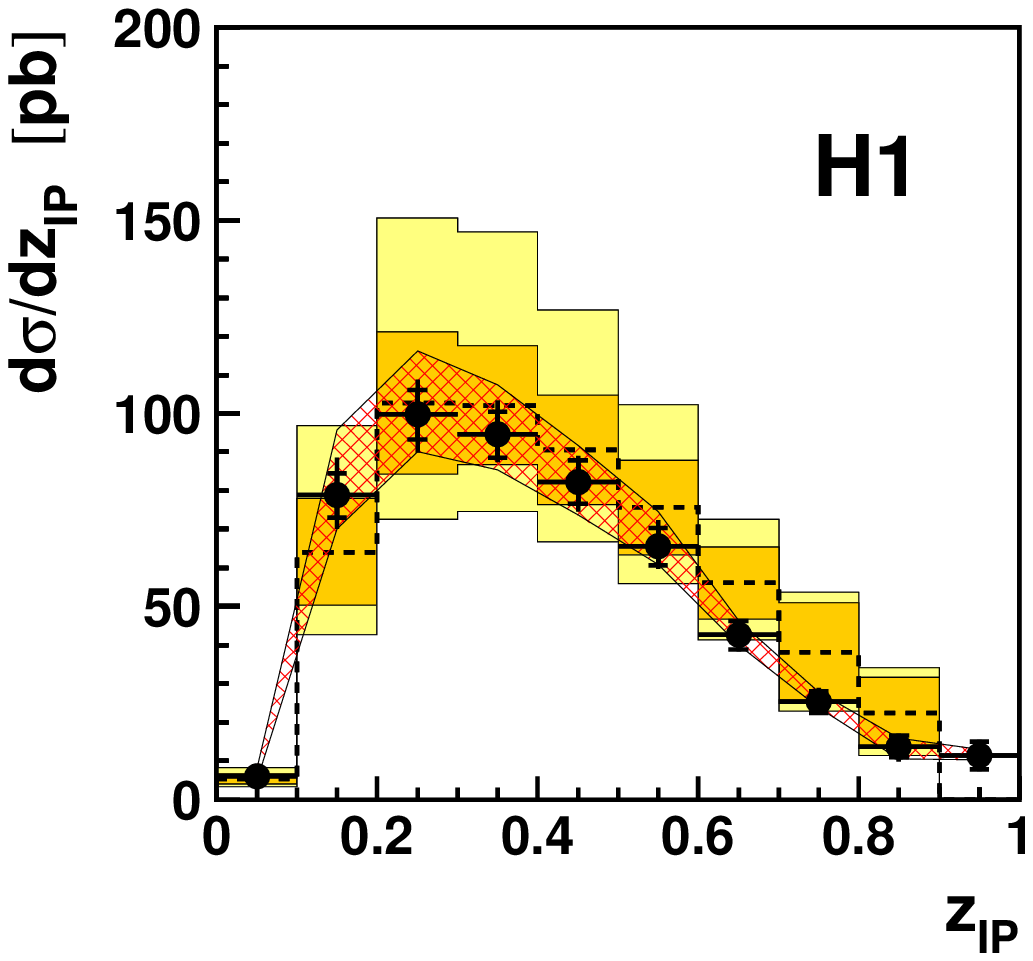}\\
 \hspace{1.cm}\includegraphics[width=0.2\textwidth]{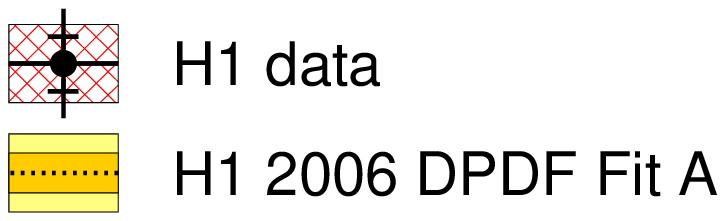} \hspace{2.5cm} \includegraphics[width=0.2\textwidth]{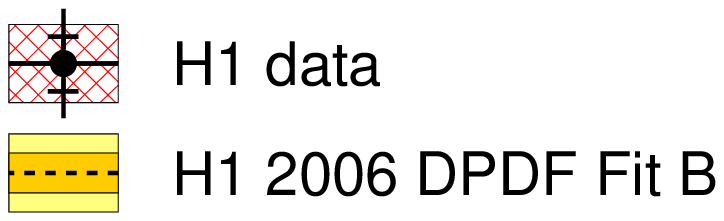} 
\caption{The differential cross section $d\sigma/dz_{I\!\!P}$ for DDIS dijet production is shown as function of $z_{I\!\!P}$.  Data (points) are compared to predictions (histograms) based on fit $A$ and $B$ explained above.} \label{fig:h1jets}
\end{figure}

One way of confirming the validity of the above approach and to differentiate between fit $A$ and $B$ is to take the parton distributions as obtained from a fit to the inclusive DDIS data and apply them to describe an exclusive channel such as DDIS dijet production.  This channel is expected to be particularly sensitive to the gluon content of the pomeron, also at high $z$.  As can be seen in Fig.\@~\ref{fig:h1jets} \cite{bib:ddisdijet_dpdf}, Fit $A$ is in good agreement with the DDIS dijet cross section at low $z_{I\!\!P}$, but overshoots the data at high $z_{I\!\!P}$.  Fit $B$, however, is in good agreement with the data at all $z_{I\!\!P}$.  This comparison therefore confirms QCD factorisation in DDIS and favours fit $B$ obtained from inclusive data.

Including the jet data in a combined fit of dijet and inclusive DDIS data yields a unique result with $\chi^2 = 196/218 {\rm\ d.o.f.}$, where both the quark and gluon distribution are constrained with similar good precision \cite{bib:ddisdijet_dpdf}.  The resulting parton densities lie close to Fit $B$ and are the most precise to date.

\section{SURVIVAL PROBABILITIES}

Although the DPDFs extracted from a fit to inclusive DDIS data from HERA can be used to predict other DDIS channels such as dijet production, they fail to describe diffractive jet production in proton-proton scattering at the TEVATRON by a factor of order 10.   This is to be expected, as QCD factorisation in not supposed to hold in proton-proton diffraction:  multi-pomeron exchanges, remnant interactions or screening may lead to additional particle production, thereby destroying the rapidity gap.  These effects can be parametrised as a rapidity gap survival probability and a lot of theoretical and experimental effort now goes to the determination of this factor.

\subsection{Survival Probability from H1 and ZEUS}

One way to study the rapidity gap survival within one experiment is provided in electron-proton diffractive photoproduction of dijets.  There one can compare interactions where the quasi-real photon interacts as a whole to interactions where the photon is resolved in a hadron-like structure so that only part of photon's momentum enters the dijet system.  Experimentally, both cases can be distinguished by reconstructing the variable $x_\gamma$:  direct photon interactions will have a reconstructed value of $x_\gamma$ close to 1, while resolved photon interactions will have lower values for $x_\gamma$.
One should note however that the separation between direct and resolved photon interactions in theoretical calculations is only possible at fixed order, as additional orders will move part of the direct photon cross section at lower order to the resolved photon cross section.

\begin{figure}[htb]
\includegraphics[width=0.28\textwidth]{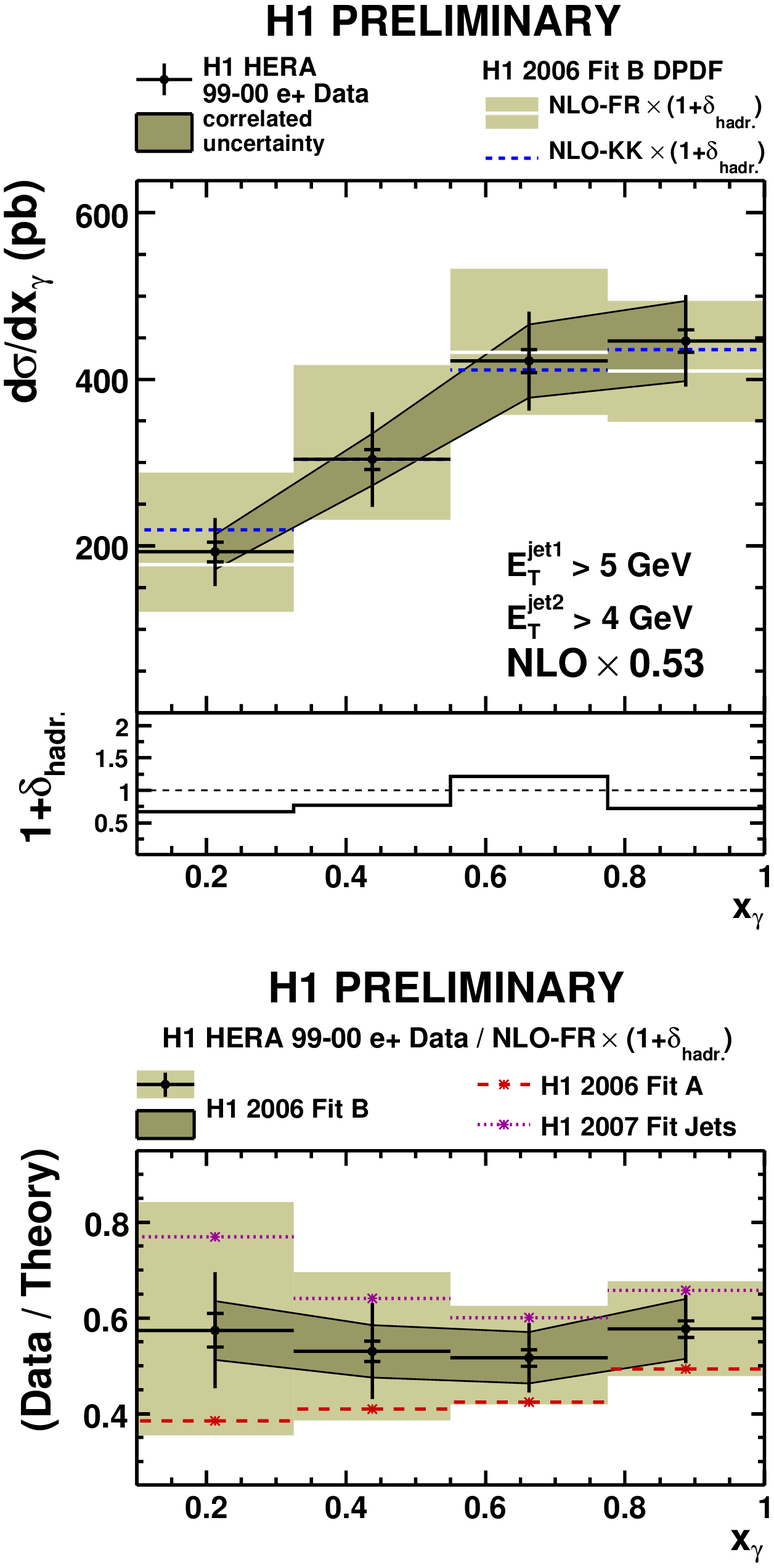}\hspace{1.5cm}\includegraphics[width=0.3\textwidth]{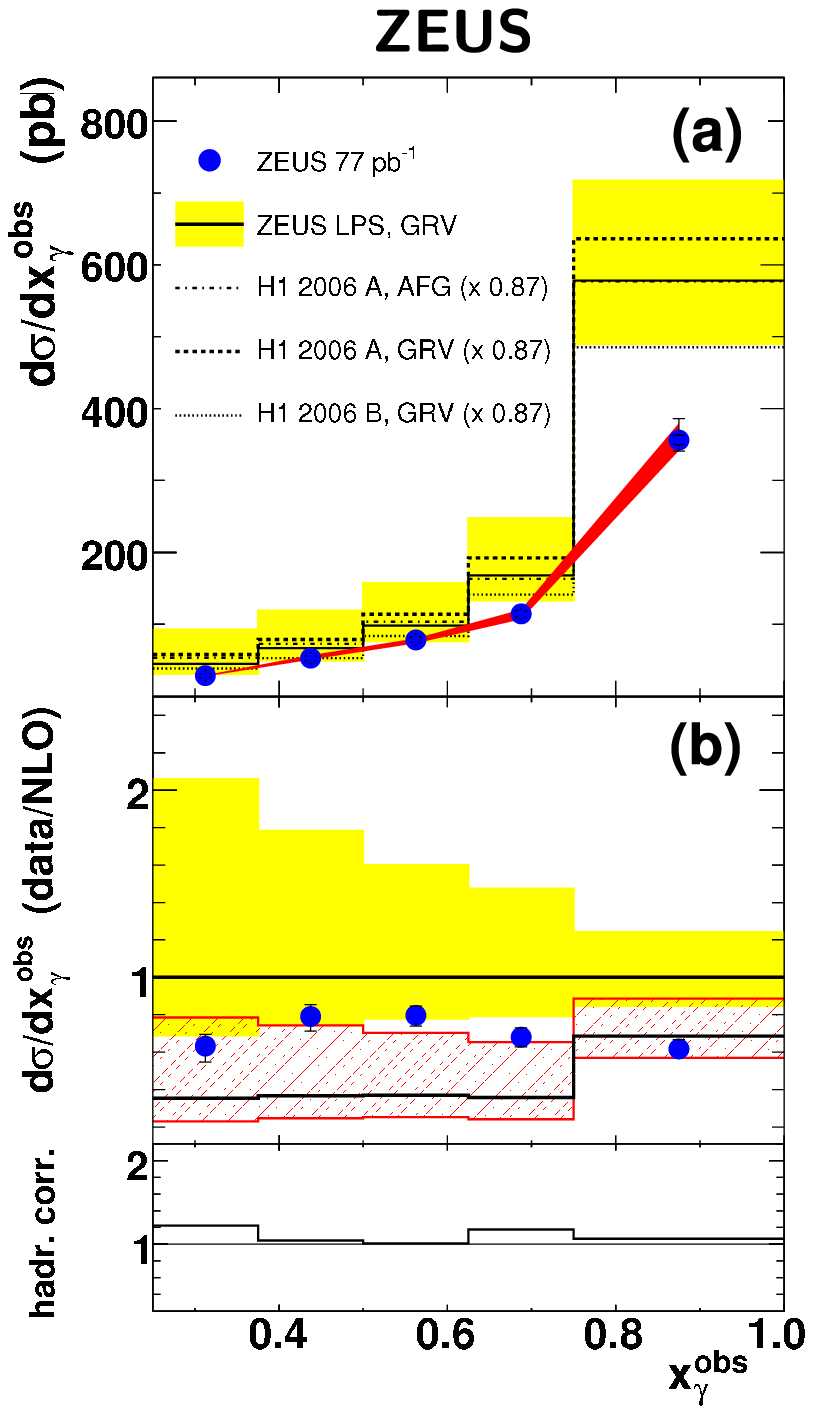}
\caption{Differential cross section and ratio of data over theory for diffractive photoproduction of dijets as function of $x_\gamma$ measured by H1 and ZEUS.} \label{fig:survprob}
\end{figure}

Both the H1 and ZEUS collaborations have studied the rapidity gap survival probability by measuring the $x_\gamma$ dependence of the cross section for diffractive dijet photoproduction \cite{bib:cerny,bib:zeusdphpjets}.  Surprisingly, although both experiments do observe a suppresion of the measured cross section when compared to the theoretical prediction without survival factor, neither experiment finds a strong dependence on $x_\gamma$ (see Fig.\@~\ref{fig:survprob}).  As a result, no evidence has been found for any difference in survival probability for interactions mediated by resolved and direct photons.  A difference in the observed survival factor between H1 and ZEUS has been traced back to different cutoffs in jet $E_T$ and a harder $E_T$ slope in data compared to NLO theory.

\subsection{Diffractive W and Z Production at the TEVATRON}

The measurement of diffractive production of vector bosons in $pp$ collisions provides another possibility to study rapidity gap survival.  Moreover, this process is also sensitive to the quark component of DPDFs.  

Using additional forward detectors available in the TEVATRON run-II (such as a miniplug calorimeter, beam shower counters and Roman pot proton taggers), the CDF collaboration obtained a measurement of the ratio of diffractive to non-diffractive $W$ and $Z$ production \cite{bib:cdfdiffW}:
\begin{eqnarray}
R^W (0.03 < \xi < 0.10, |t| < 1 {\rm\ GeV}^2) = [0.97 \pm 0.05 {\rm\ (stat.)} \pm 0.11 {\rm\ (syst.)}]\% \\
R^Z (0.03 < \xi < 0.10, |t| < 1 {\rm\ GeV}^2) = [0.85 \pm 0.20 {\rm\ (stat.)} \pm 0..11 {\rm\ (syst.)}]\%
\end{eqnarray}
These results are in good agreement with previous Run-I results of D0 and CDF obtained with the rapidity gap method.

\section{CENTRAL EXCLUSIVE PRODUCTION AT THE TEVATRON}

Central exclusive production in $pp$ collisions is a particularly interesting channel for the discovery or study of the Higgs (see below).  Theoretical calculations of this process however suffer from the large uncertainty that exists on the rapidity gap survival factor.  It is therefore of utmost importance to establish central exclusive production of a variety of known final states, so that these can be used as ``standard candles'' in the search or study of Higgs particles through the CEP channel.

\subsection{Central Exclusive Production of Dijets}

\begin{figure}[htb]
\includegraphics[width=0.4\textwidth]{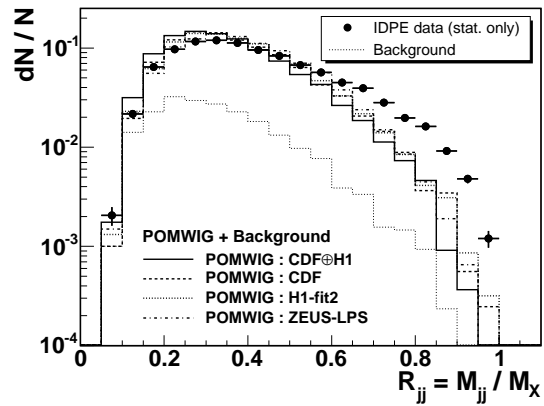}
\caption{The $R_{jj}$ distribution observed in DPE data (points) is compared to predictions by POMWIG (histograms) based on different DPDFs extracted from data.} \label{fig:cepdijets}
\end{figure}

The CDF collaboration searched for CEP of dijets by looking for an excess in the distribution of the dijet mass fraction $R_{jj} = \frac{M_{jj}}{M_X}$ in DPE events \cite{bib:cdfcepdijets}.  Events where dijets are produced exclusively should show up at $R_{jj} \approx 1$.  In Fig.\@~\ref{fig:cepdijets} the observed $R_{jj}$ distribution is compared to the POMWIG Monte Carlo model.  This model uses DPDFs extracted from data as input but does not include exclusive production of dijets. An excess of data over the POMWIG prediction is observed at high $R_{jj}$, indicating that exclusive dijet events are present in the data.  As a cross-check, a similar search was made for an excess of $b$-tagged jets.  Such an excess was not found, as is expected due to spin selection rules.

\begin{figure}[htb]
\includegraphics[width=0.4\textwidth]{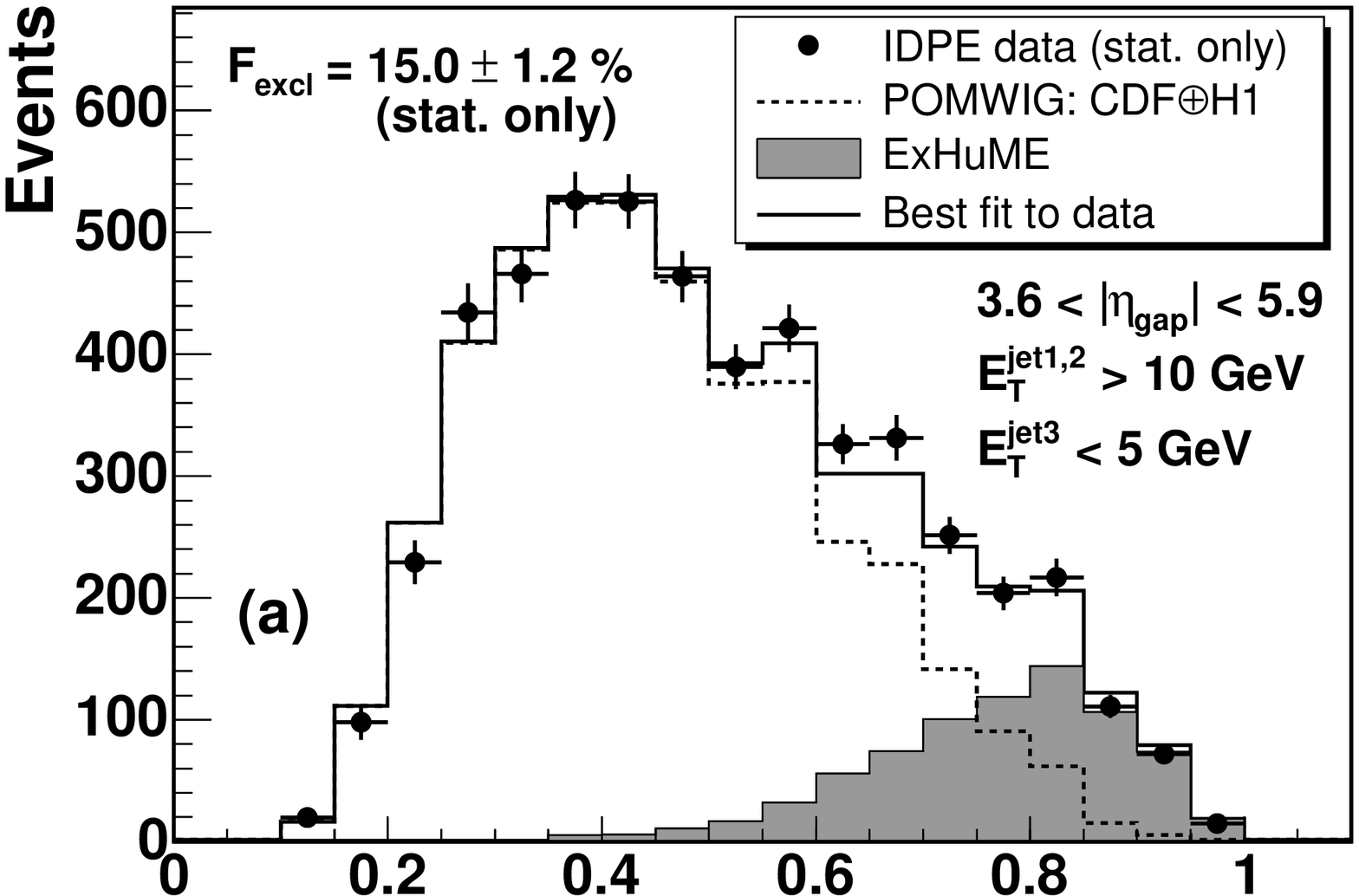}\hspace{1.5cm}\includegraphics[width=0.4\textwidth]{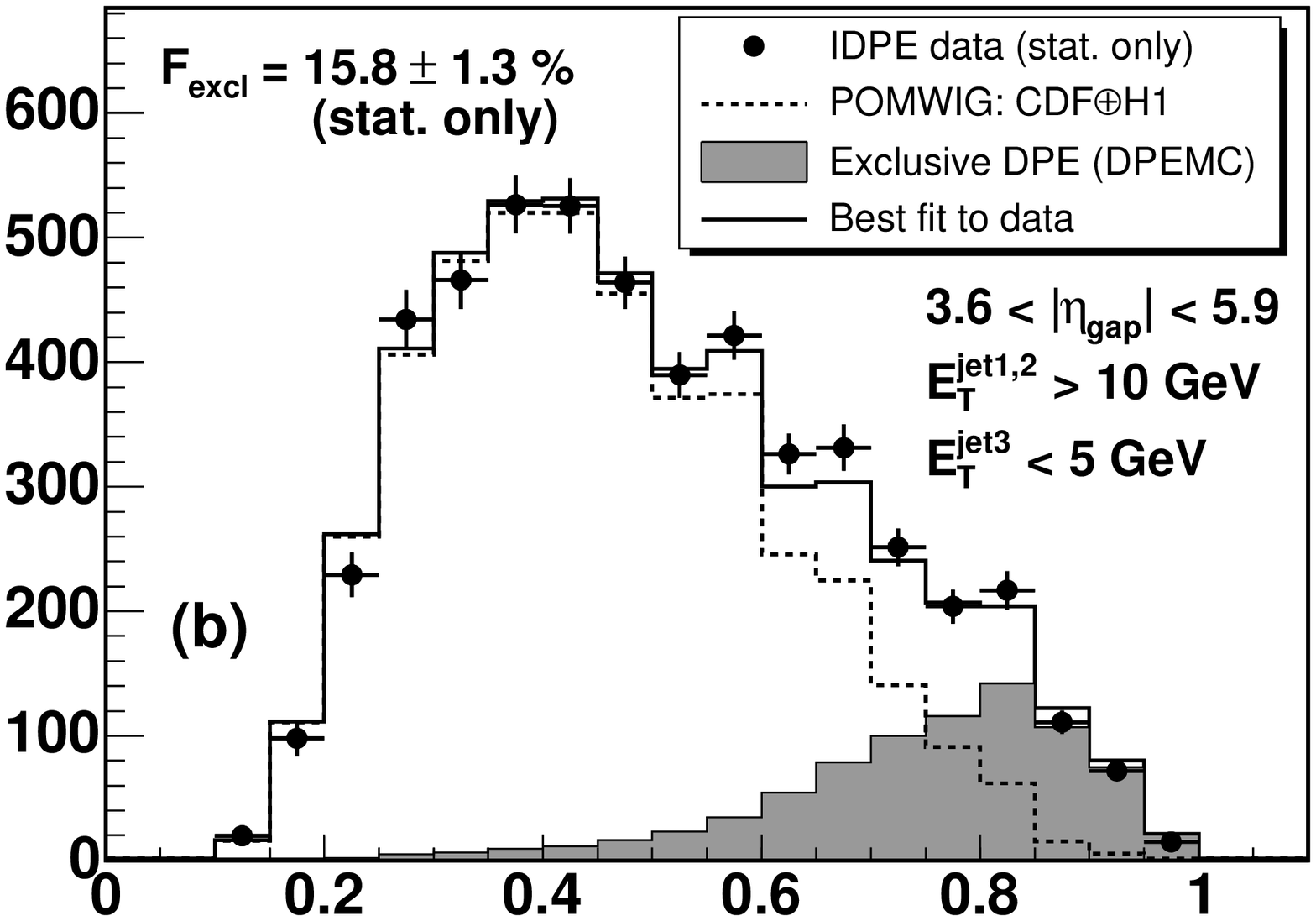}
\caption{Fits to the data (points) distribution of $R_{jj}$ using the sum of POMWIG and CEP models (histograms).  The normalisation of the CEP models is left free, yielding a fraction of exclusive events around 15\% in both cases.} \label{fig:exhumedpemc}
\end{figure}

\begin{figure}[htb]
\includegraphics[width=0.35\textwidth]{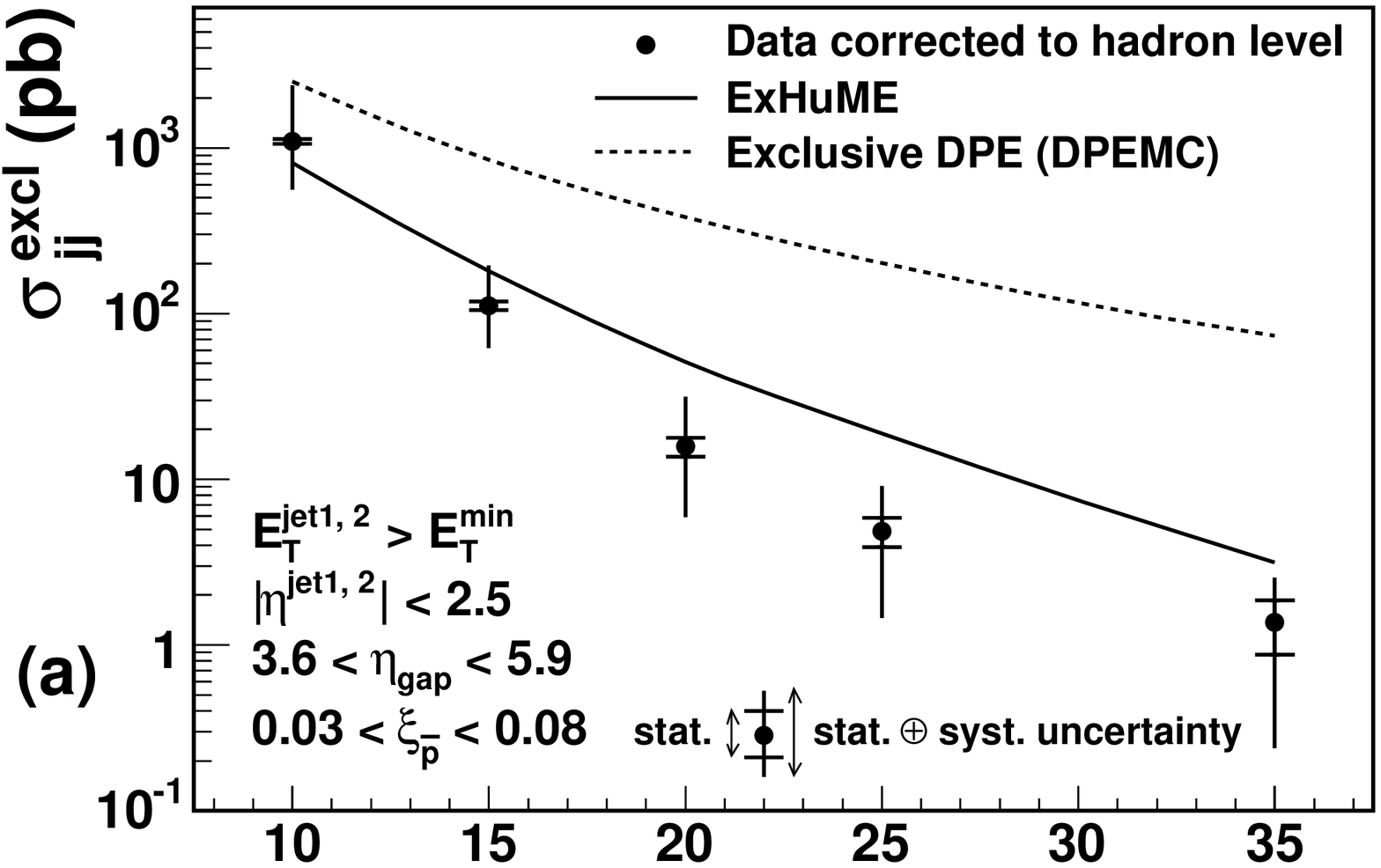}\hspace{1.5cm}\includegraphics[width=0.35\textwidth]{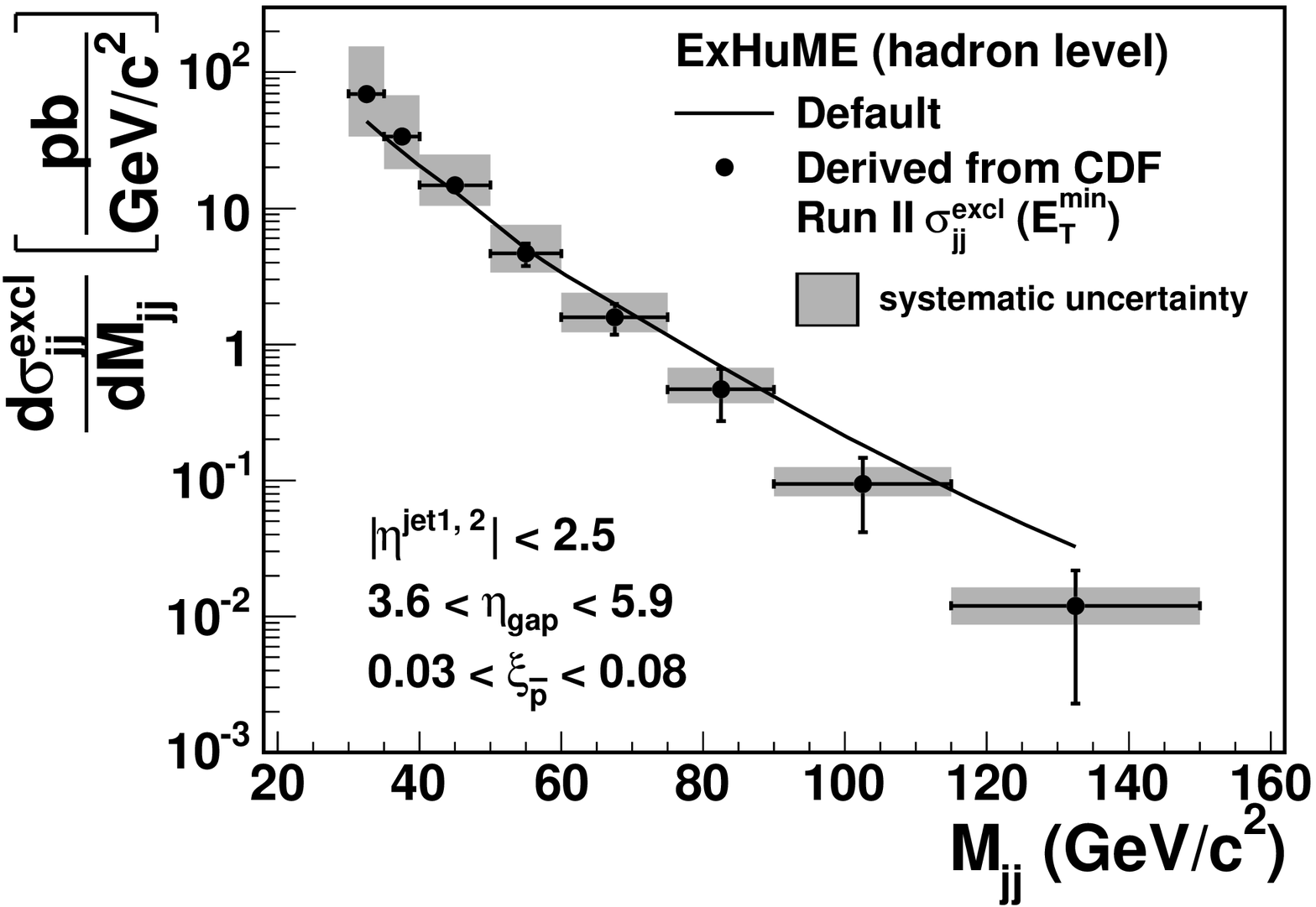}
\caption{(left) Cross section for CEP of dijets as function of transverse energy of the lowest $E_T$ jet. (right) Differential cross section for CEP of dijets as function of dijet invariant mass, extracted from data using ExHuME.} \label{fig:jetetmjj}
\end{figure}

After applying further selections to enhance the exclusive signal, a fit to the data distribution of $R_{jj}$ was made using the sum of POMWIG and specific models for CEP of dijets with a free normalisation of the CEP models (see Fig.\@~\ref{fig:exhumedpemc}).  Two models have been used: ExHuME \cite{bib:exhume}, which is based on a LO pQCD calculation \cite{bib:khoze}, and DPEMC \cite{bib:dpemc}, which is an exclusive DPE Monte Carlo model based on Regge Theory \cite{bib:bialas}.  Both models are able to describe the excess at high $R_{jj}$ well.  However, when looking at the jet $E_T$ distribution the ExHuME model is favoured.  This model also describes the $M_{jj}$ distribution reasonable well (see Fig.\@~\ref{fig:jetetmjj}).

\subsection{Central Exclusive Production of Diphotons and Dileptons}

Other CEP final states have also been investigated by the CDF Collaboration.  In a sample of $532 {\rm\ pb}^{-1}$ of Run-II data, 3 exclusive diphoton events were found with $E_T^\gamma > 5 {\rm\ GeV}$ and $|\eta^\gamma| < 1$ \cite{bib:diphotons}. Two of the candidate events are almost certainly diphoton final states, although the $\pi^0\pi^0$ or $\eta\eta$ hypotheses cannot be completely discounted.  The probability that other processes fluctuate to 3 events or more is $1.7 \times 10^{-4}$.  The kinematics of the events found in data are in agreement with the ExHuME model, which predicts $0.8^{+1.6}_{-0.5}$ events.  The upper limit for the CEP diphoton cross section has been set at 410 fb (95 \% CL).

Exclusive production of dileptons can occur though two-photon exchange, a nearly pure QED process.  Using the same dataset as above, CDF found 16 candidate events with $E_T^e > 5 {\rm\ GeV}$ and $|\eta^e| < 2$, over an expected background of $1.9 \pm 0.3$ \cite{bib:dileptons}.  The measured cross section is $1.6^{+0.5}_{-0.3} {\rm\ (stat.)} \pm 0.3 {\rm\ (syst.)} {\rm\ pb}$, which agrees with theoretical expectations.

\section{FORWARD LOOK TO THE LHC}

\subsection{Diffractive W Production}

CMS has studied the feasibility of observing single diffractive $W$ production in $100 {\rm\ pb}^{-1}$ of LHC data \cite{bib:cmsdiffW}. The diffractive selection is based hadron activity measured in the forward calorimeters HF and CASTOR, as well as particle multiplicity detected in the central tracker.  Especially the CASTOR calorimeter with an acceptance of $5.2 < \eta < 6.6$ is essential to achieve a signal-to-background ratio of up to 20.  For a rapidity gap survival factor $S^2 = 0.05$, $O(100)$ reconstructed signal events are expected.

\subsection{Exclusive Dilepton Production and $\boldsymbol{\Upsilon}$ Photoproduction}

Exclusive dilepton production, $pp \to pp l^+l^-$, through double photon exchange, is a nearly pure QED process and can therefore be used for luminosity monitoring with a precision of down to 4\%.  The measurement of this process can also help in the study of lepton identification in the main CMS detector and for the calibration of forward proton detectors.  

The CMS collaboration prepares the measurement of exclusive dilepton production based on the detection of centrally produced $e^+e^-$ or $\mu^+\mu^-$ pairs \cite{bib:ovyn}.  The main uncertainty in this analysis will be due to the inelastic background where one of the protons dissociates.  Again, the use of the forward calorimeters (CASTOR and ZDC) can greatly reduce this background.

\begin{figure}[htb]
\includegraphics[width=0.62\columnwidth]{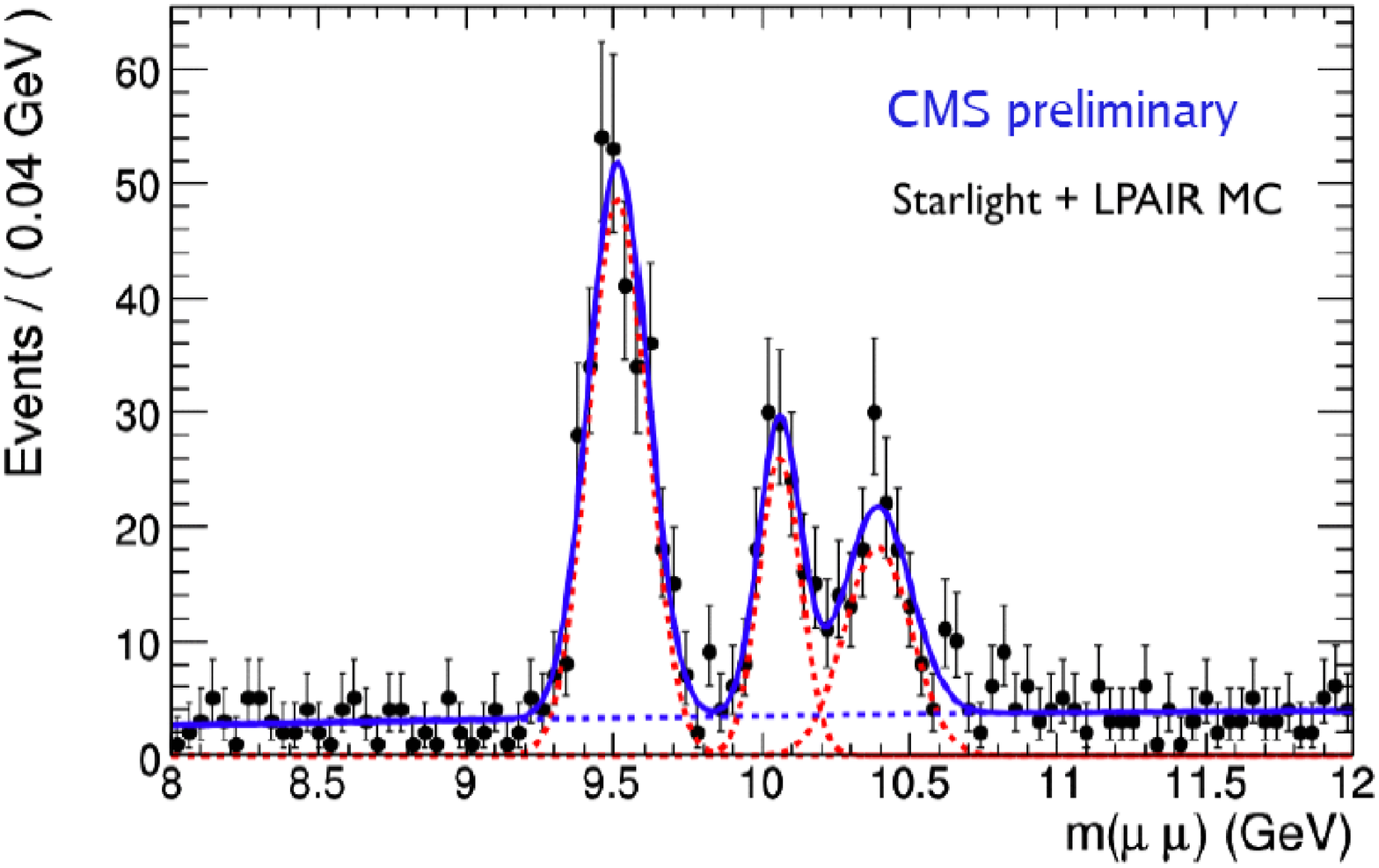}\includegraphics[width=0.38\columnwidth]{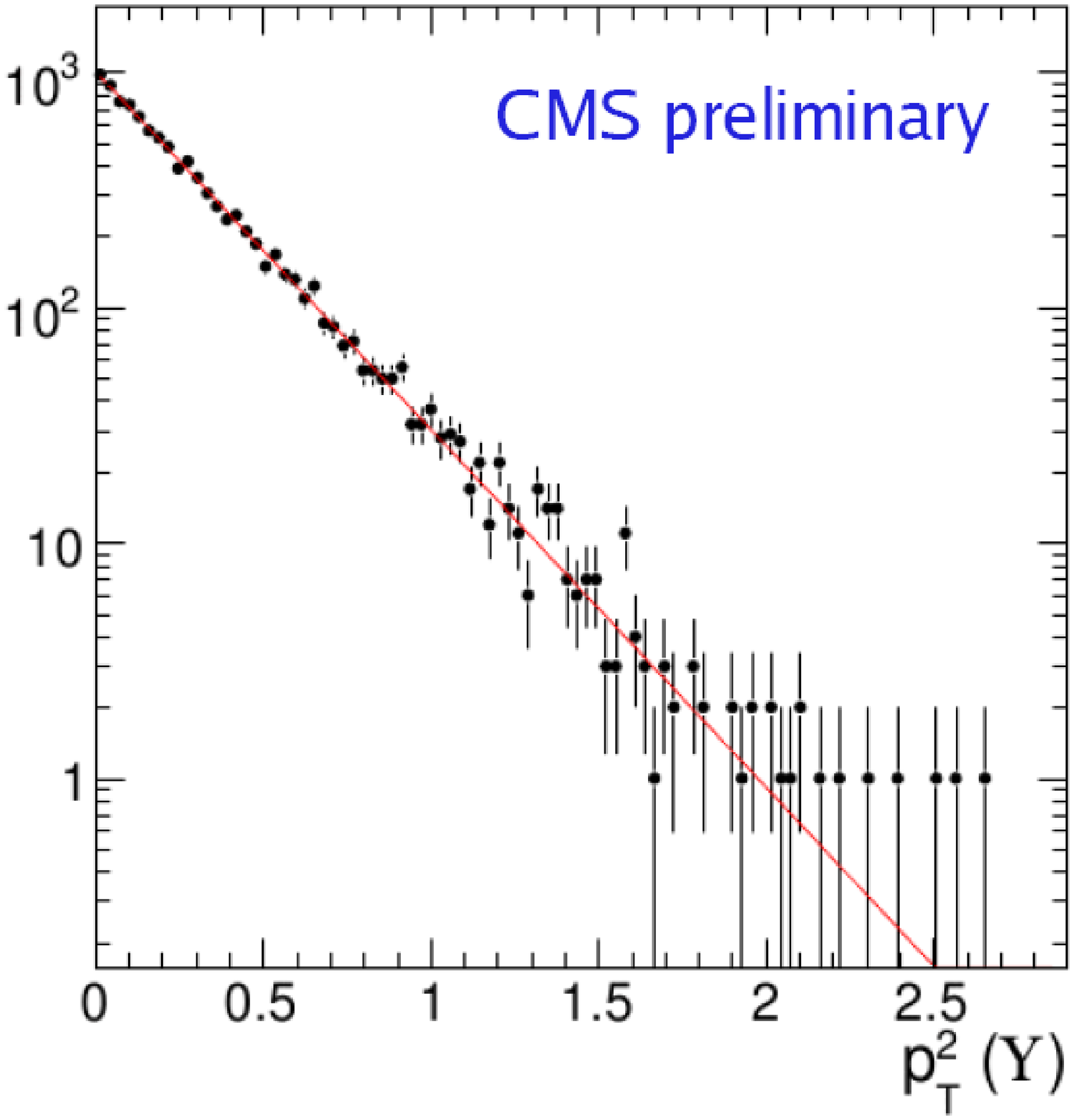}
\caption{(left) Invariant mass spectrum for diffractive $\Upsilon$ photoproduction $pp \to pp\Upsilon, \Upsilon \to \mu^+\mu^-$; (right)  $p_T^\Upsilon$ distribution.}
\label{fig:upsilon}
\end{figure}

The $p_T$ threshold used for the detection of muon pairs is low enough to allow the reconstruction of the $\Upsilon$ mass peaks (see Fig.\@~\ref{fig:upsilon}).  Here, the $\Upsilon$ is produced in diffractive photoproduction processes.  The analysis of the process therefore allows to constrain the gluon distribution in the proton at low Bjorken-$x$ and to study diffractive and QCD models.   A preliminary CMS analysis shows that the resolution is good enough to resolve different $\Upsilon$ resonances and to extract the exponential slope parameter $b$ from the $p_T$ spectrum.

\subsection{Central Exclusive Higgs Production}

The central exclusive production of Higgs particles has some advantages over inclusive channels: QCD $b\bar{b}$ backgrounds are suppressed due to the $J_z = 0$ spin selection rule, an accurate determination of the Higgs mass is possible through the measurement of the outgoing proton momenta and azimuthal angular correlations may shed information on the spin-parity of the Higgs.  

Given the large uncertainty on the rapidity gap survival factor, a data-driven calibration is however mandatory.  Here the observation of central exclusive producion of dijets, diphoton, $\chi_c$ particles, etc.\@ may serve to calibrate models.  The calculation in \cite{bib:khoze} predicts a CEP standard model Higgs cross section of 3 fb at the LHC.

In particular scenarios of the MSSM and NMSSM, CEP may be the most probable channel for a discovery \cite{bib:cephiggs}.

\section{SUMMARY}

HERA measurements of inclusive diffractive deep-inelastic scattering now give consistent results for all methods and experiments.  Diffractive parton density functions are extracted; the combined analysis of inclusive and dijet diffractive deep-inelastic data samples obtained by H1 provides to most precise parton densities to date.  

No suppression of the cross section for diffractive photoproduction of dijets is observed  for resolved photons.  The survival probability does seem to increase however for higher $E_T$ jets.

Central exclusive dijet production is observed at the TEVATRON and can serve as a ``standard candle'' process for the study of central exclusive production of Higgs particles at the LHC.  Theoretical models based on LO pQCD calculations are in agreement with this data.

Plans to establish diffractive signals at the LHC are being developed.

Central exclusive production of Higgs particles has some advantages over inclusive channels and can potentially lead to a discovery in some MSSM and NMSSM Higgs scenarios.

% If you have acknowledgments, this puts in the proper section head.
\begin{acknowledgments}
I would like to thank the organisers of HCP 2008 for their hospitality and the invitation to Galena, Illinois.  I appreciate the hard work of all members of the H1, ZEUS, CDF and CMS Collaborations who contributed to the results presented here, by collecting and analysing the experimental data.
\end{acknowledgments}

\end{document}